\documentclass[aps,prd,twocolumn,twoside,preprintnumbers,superscriptaddress,floatfix]{revtex4-2}

\usepackage{epsfig}
\usepackage{amsmath}
\usepackage{amssymb}
\usepackage{latexsym}
\usepackage{xspace}
\usepackage[colorlinks=true,linktocpage=true,linkcolor=blue,citecolor=blue,allcolors=blue]{hyperref}
\usepackage{url}
\usepackage[utf8]{inputenc}
\usepackage{indentfirst}
\usepackage{enumerate}
\usepackage{color}
\usepackage{relsize}
\usepackage{graphicx}
\usepackage{dcolumn}
\usepackage{bm,bbold}
\usepackage[caption=false,position=top]{subfig}
\usepackage[english]{babel}
\usepackage{slashed}
\usepackage[normalem]{ulem}

\newcommand{\be}{\begin{equation}}
\newcommand{\ee}{\end{equation}}
\newcommand{\bea}{\begin{eqnarray}}
\newcommand{\eea}{\end{eqnarray}}

\begin{document}

\title{Dark gauge boson emission from supernova pions}

\author{Chang Sub Shin}
 \email{csshin@cnu.ac.kr}
 \affiliation{Department of Physics and Institute of Quantum Systems (IQS), \\ Chungnam National University, Daejeon 34134, Republic of Korea}
 \affiliation{Center for Theoretical Physics of the Universe, Institute for Basic Science (IBS), Daejeon, 34126, Korea}
 \affiliation{Korea Institute for Advanced Study, Seoul 02455,  South Korea}
 
 \author{Seokhoon Yun}
 \email{seokhoon.yun@pd.infn.it}
\affiliation{Center for Theoretical Physics of the Universe, Institute for Basic Science (IBS), Daejeon, 34126, Korea}
\affiliation{Dipartimento di Fisica e Astronomia, Universit\`a degli Studi di Padova, Via Marzolo 8, 35131 Padova, Italy}
\affiliation{INFN, Sezione di Padova, Via Marzolo 8, 35131 Padova, Italy}

\date{\today}

\begin{abstract}

The hot, neutron-rich, and dense circumstance in core-collapse supernovae provides a source of negatively charged pions that may make up a significant portion of the matter.
These abundant thermal pions offer an opportunity to populate light and hidden hypothetical particles.
In this study, we discuss the dark gauge boson production via reactions involving supernova pions. The rate of this production is determined by the isovector nucleon coupling. 
We consider two toy models, the dark photon and the gauged $B-L$ models, that carry the typical distinct isovector nucleon coupling structure in the medium.
Pion-induced dark gauge bosons leave an imprint on several observational consequences associated with supernova.
Their sizable emissivity and characteristic hard spectral distribution result in the stringent constraints on the dark gauge boson models, in particular at masses above the two electron mass.

\end{abstract}

\maketitle


\section{Introduction}

A star with the initial mass greater than about 8 solar mass develops an iron core where nuclear fusion ceases to take place.
As the iron core accumulates mass beyond the Chandrasekhar limit~\cite{Chandrasekhar:1931ftj,Chandrasekhar:1935zz}, it eventually undergoes gravitational collapse, resulting in the formation of a shock wave at the surface of the core (an internal object with a stiff equation of state is called the protoneutron star) and subsequently blows up the outer layers with the help of the delayed partial neutrino energy conduction~\cite{Bethe:1985sox,Kotake:2005zn,Burrows:2012ew,Janka:2012wk}.
This extraordinary and explosive event is referred to as a type-II supernova.

One notable example of a core-collapse supernova is SN1987A, which occurred on February 23, 1987, in the Large Magellanic Cloud. It stands as the only event in which an energetic burst of neutrinos was directly detected.
Apart from the ejection of the outer part, a compact object associated with the core may remain in the center.
The recent observations of the heated dust clumps at the inferred position (with a kick velocity)  from the Atacama Large Millimeter Array~\cite{Cigan:2019shp,Page:2020gsx} and the analysis of hard X-ray spectra performed by {\it Chandra}, {\it XMM-Newton}, and {\it NuSTAR}~\cite{Greco:2021sfi,Greco:2022cto} strongly suggest the presence of a neutron star as the promising remnant of SN1987A.

Supernovae provide a valuable testing ground~\cite{Raffelt:1996wa} for exploring new light particles beyond the Standard Model (SM) such as axions, dark photons, and other similar particles that are  challenging to detect in terrestrial experiments. Indeed, novel particles can be copiously produced in supernova, particularly inside the protoneutron star (pNS), through particle reactions occurring in its dense medium.
The interior of the pNS is characterized by high temperature ($T\sim 30\,{\rm MeV}$ for a few seconds after birth) and extreme density ($\rho \gtrsim \rho_{\rm sat} \simeq 2.6 \times 10^{14}\,{\rm g}\,{\rm cm}^{-3}$).
In such a neutron-rich environment ($n_p/n_B \sim 20\,\%$ with $n_p$ and $n_B$ the proton and baryon number density, respectively), the production of these new particles is primarily influenced by nucleon-nucleon bremsstrahlung mediated by the strong interactions, which has been extensively studied in Refs.~\cite{Brinkmann:1988vi,Raffelt:1991pw,Iwamoto:1992jp,Rrapaj:2015wgs,Fischer:2016cyd,Shin:2021bvz,Balaji:2022noj}.

Recent studies have put forward the idea that, in addition to electrons, other negatively charged particles like  muons~\cite{Bollig:2017lki} and pions~\cite{Fore:2019wib} might constitute a significant fraction ($\mathcal{O}(1)\,\%$ ratio to $n_B$) in the pNS. The non-negligible abundance of these particles can be attributed to the high temperature and the hight electron chemical potential, which is determined by the difference between the nucleon chemical potentials ($\mu_n - \mu_p$). For more details, please see Refs.~\cite{Bollig:2017lki,Fore:2019wib} and the references therein. The inclusion of thermally equilibrated muons and pions has implications for various aspects of core-collapse supernovae, such as their potential to  facilitate neutrino-driven supernova explosions~\cite{Bollig:2017lki}, the equation of state of the pNS, or the reduced mean free path of low energy muon neutrinos~\cite{Fore:2019wib}.
In addition, the scattering of negatively charged particles can make a significant contribution to the production of new particles~\cite{Carenza:2020cis,Fischer:2021jfm,Choi:2021ign,Lella:2022uwi}.

In this study, our focus is on investigating the production of feebly interacting vector bosons associated with a novel Abelian gauge symmetry within supernovae. We specifically examine the mechanism of their production through thermal pion scatterings which exhibit similarities to the Panofsky process~\cite{Panofsky:1950he}. We also explore the implications of these vector bosons on various supernova observations.
These light vector bosons are often referred to as a dark gauge bosons~\cite{Fayet:1980ad,Fayet:1980rr,Fayet:1990wx,Fayet:2016nyc,Arkani-Hamed:2008hhe} denoted by $\gamma^\prime$ or $A^\prime_\mu$.
In comparison to the more commonly considered nucleon-nucleon bremsstrahlung~\cite{Shin:2021bvz}, 
pionic processes yield dark gauge bosons with a comparable emission rate, but  they possess an one order of magnitude larger energy due to the thermal pion energy threshold (including its mass), similar to the case of axions~\cite{Carenza:2020cis,Fischer:2021jfm,Choi:2021ign}.  The significant emissivity and harder energy spectrum of these bosons have remarkable implications for observations associated with supernovae such as the duration of the observed signal from SN1987A, observations of supernova explosion energy, the estimated injection rate of positrons in the galaxy as a source of diffuse gamma-ray emission, and the absence of prompt gamma-rays in SN1987A. 
These implications eventually lead to stringent constraints on models involving dark gauge bosons, particularly for masses above two times the electron mass, as they can decay into electron-positron pairs when coupled to the electron current. 

The paper is organized as follows.
In Sec.~\ref{sec:effective}, we describe the effective couplings of dark gauge bosons to the SM particles in a dense medium with the proper treatment of the effective field theory formalisms and thermal effects.
In Sec.~\ref{sec:production}, we investigate the dark gauge boson production from negatively charged thermal pions in the pNS.
Sec.~\ref{sec:implications}  is dedicated to discussing the phenomenological implications of two specific extensions of the SM with gauged $U(1)$: the dark photon model (only involving the kinetic mixing~\cite{Holdom:1985ag}) and the gauged $B-L$ model (which is anomaly free if right-handed neutrinos are introduced).
We then provide discussions and conclusion in Sec.~\ref{sec:conclusions}.

\section{Effective  Lagrangian in medium}
\label{sec:effective}

The most general effective Lagrangian with a gauged $U(1)$ extension of SM at scales above the strong confinement ($\mu \sim  2\,{\rm GeV}$) is expressed as follows
\be
\begin{split}
\mathcal{L} = & -\frac{1}{4}F_{\mu\nu}F^{\mu\nu} -\frac{1}{4} F^\prime_{\mu\nu}F^{\prime \mu\nu} + \frac{\varepsilon}{2}F_{\mu\nu}F^{\prime \mu\nu}  \\
& + \frac{m_{\gamma^\prime}^2}{2}A_\mu^\prime A^{\prime \mu} + e A_\mu J_{\rm EM}^\mu + g^\prime A_\mu^\prime J^{\prime \mu} \, ,
\end{split}
\ee
where $F_{\mu\nu} = \partial_\mu A_\nu - \partial_\nu  A_\mu \left( F_{\mu\nu}^\prime= \partial_\mu A_\nu^\prime - \partial_\nu  A_\mu^\prime\right)$ is the field strength of the photon $A_\mu$ (the dark gauge boson $A^\prime_\mu$ with its mass  $m_{\gamma^\prime}$), $\varepsilon$ is the dimensionless kinetic mixing~\cite{Holdom:1985ag} between the photon and dark gauge boson fields, and $eJ_{\rm EM}^\mu$ and $g^\prime J^{\prime \mu}$ denote the electromagnetic and dark $U(1)$ currents, respectively.
Although $J^{\prime \mu}$ involves in general both the vector and axial-vector currents, we focus on the case of vector gauge charge assignment to the SM fields, which is written by
\be
g^\prime J^{\prime \mu} = \sum_{f=q,l} g^\prime q_f^\prime \bar{f}\gamma^\mu f \, 
\label{eq:DarkCurrent}
\ee
with the dark $U(1)$ quantum number $q^\prime_f$ of the particle $f$ for the lighter quarks $(q = u,d,s)$ and the leptons $(l = e,\mu)$.

After the strong confinement, the Lagrangian within the perturbative theory framework is unable to describe the physics in the strong sector and one must rely on non-perturbative techniques.
The formalism of Chiral Perturbation Theory (ChPT)~\cite{Weinberg:1978kz,Gasser:1983yg,Gasser:1984gg} accounts for the effective couplings of hadronic resonances  mediated by quark interactions.
The basic approach is to match terms that are identically transformed under the underlying symmetries such as the chiral symmetry, isospin symmetry, charge conjugation, etc.
Focusing on the relevant interactions of nucleons and pions, the $SU(2)\times SU(2)$ symmetric effective Lagrangian corresponds to
\be
\begin{split}
\mathcal{L}_{\rm ChPT} = & \frac{f_\pi^2}{4}\, {\rm Tr}\left[D^\mu U^\dagger D_\mu U\right]  \\
& + \bar{N} i \slashed{D}N + \frac{g_A}{2} \bar{N} \gamma^\mu \gamma^5 u_\mu N \, ,
\end{split}
\label{eq:ChPT}
\ee
where $f_\pi \simeq 93\,{\rm MeV}$, $g_A=1.2723(23)$, $N = (p\, ,n)^{\rm T}$ is the nucleon doublet, and 
\be
\begin{split}
U = \exp \left[i\frac{\pi^a \sigma^a}{f_\pi }\right] = u^2  \, ,  \quad  u_\mu \equiv  -i u^\dagger \left(D_\mu U\right) u^\dagger 
\end{split}
\ee
with the Pauli matrices  $\sigma^a$ for the pion representation 
\be
\pi^a \sigma^a = \left(
\begin{tabular}{cc}
$\pi^0$ & $\sqrt{2}\pi^+$ \\
$\sqrt{2}\pi^-$ & $-\pi^0$
\end{tabular}
\right) \, .
\ee
The covariant derivatives of the pion and nucleon fields with respect to a gauge field of our interest 
\bea
v_\mu = (c^0  \mathbb{1}  + c^a \sigma^a) V_\mu
\label{eq:vmu}
\eea
that couples to the vector currents of the up ($\bar{u}\gamma^\mu u$) and down ($\bar{d}\gamma^\mu d$) quarks, are defined as~\cite{Scherer:2002tk}
\bea
D_\mu U & = & \partial_\mu U - i v_\mu U + i U v_\mu \, , \\
D_\mu N & = & \left( \partial_\mu + \Gamma_\mu - i  v^{(s)}_\mu \right) N \, ,
\eea
with
\bea
\Gamma_\mu & = & \frac{1}{2}\left[u\left(\partial_\mu - i v_\mu\right)u^\dagger + u^\dagger\left(\partial_\mu - i v_\mu\right)u \right] \, , \\
v^{(s)}_\mu & = & {\rm Tr}\left[v_\mu\right] \mathbb{1} \, .
\eea
One can straightforwardly check the pion-nucleon couplings from the second term in the second line of Eq.~\eqref{eq:ChPT}
\be
\begin{split}
\frac{g_A}{2 f_\pi} & \left[\partial_\mu \pi^0 \left(\bar{p}\gamma^\mu\gamma^5 p - \bar{n}\gamma^\mu\gamma^5 n\right)\right. \\
& \left. +\sqrt{2} \partial_\mu \pi^+ \bar{p}\gamma^\mu \gamma^5 n + \sqrt{2} \partial_\mu \pi^- \bar{n}\gamma^\mu \gamma^5 p \right] \, .
\end{split}
\ee
The leading order hadronic interactions of $V_\mu$, which are of relevance for our purpose, are provided as follows
\be
\begin{split}
&  \hskip -0.1cm  V_\mu \Big( \sum_{N=p,n} c_N \bar{N}\gamma^\mu N + i \left(c_p-c_n\right) \Big[\pi^- \overleftrightarrow{\partial^\mu} \pi^+ \Big]  \Big.  \\
&  \Big. \hskip 0.1cm+ i\left(c_p-c_n\right)  \frac{g_A}{\sqrt{2}f_\pi}\left[\pi^- \bar{n}\gamma^\mu \gamma^5 p - \pi^+\bar{p}\gamma^\mu \gamma^5 n\right] \Big)\, ,
\end{split}
\label{eq:VInt}
\ee
where $\pi^- \overleftrightarrow{\partial^\mu} \pi^+ =\pi^- \left(\partial_\mu \pi^+\right) - \left(\partial_\mu \pi^-\right)\pi^+$.
Because we are considering the coupling of the photon and the dark gauge boson, we take $c^1=c^2=0$ in Eq.~(\ref{eq:vmu}). 
Eq.~\eqref{eq:VInt} incorporates the expected couplings to the photon $A_\mu$ 
\be
c_p(A_\mu) =  e  \, , \quad c_n(A_\mu)  =0 \, ,
\ee
and to the dark gauge boson $A_\mu^\prime$
\bea
c_p(A_\mu^\prime)  =  g^\prime (2q_u^\prime + q_d^\prime) \, , \quad  c_n(A_\mu^\prime)  =g^\prime (q_u^\prime + 2 q_d^\prime) \, .
\eea
For convenience, the dark $U(1)$ charges of the nucleon are denoted by
\be
q_p^\prime = 2q_u^\prime + q_d^\prime\, , \qquad  q_n^\prime = q_u^\prime + 2 q_d^\prime \, .
\ee

The effective couplings between dark gauge bosons and hadrons, as well as leptons, in a hot and dense plasma exhibit notable differences and peculiarities compared to the vacuum. 
The thermal field theory provides a rigorous framework to understand the behavior of these effective couplings within a medium, taking into consideration the external momentum of the dark gauge bosons 
\be 
p^{\mu}_{\gamma^\prime} = \big(\omega, \vec{k} \big) \, .
\ee
Interactions with electrons, being the lightest charged particles, play a dominant role in the in-medium effect~\cite{Rrapaj:2015wgs}.
In the regime of weak coupling, which governs the linear response, the effective couplings of dark gauge bosons follow th same form of operators in Eq.~\eqref{eq:DarkCurrent} for leptons, Eq.~\eqref{eq:VInt} for hadrons, with the following redefined couplings (for details, see Refs.~\cite{Rrapaj:2015wgs,Hong:2020bxo,Shin:2021bvz})
\be
\begin{split}
g^\prime q_f^\prime \, \rightarrow \, g^\prime \tilde{q}_{f}^\prime  = & \,  g^\prime \left(q_e^\prime q_f - q_e q_f^\prime \right) \\
& + \left(\varepsilon e - g^\prime q_e^\prime\right) q_f\frac{m_{\gamma^\prime}^2}{m_{\gamma^\prime}^2-\pi_{\rm T,L}} \, ,
\end{split}
\label{eq:VchargeinM}
\ee
where $\tilde{q}^\prime_f$ denotes an effective dark $U(1)$ charge in medium for $f=\left(l,p,n\right)$.
The $\varepsilon$ dependence in Eq.~(\ref{eq:VchargeinM}) reflects the canonical diagonalization due to the kinetic mixing.  
The $\pi_{\rm T,L}$ quantify the refractive properties of the intermediate photon propagator in the medium for the transverse and longitudinal polarizations of the external dark gauge boson, respectively. Their real parts are given by~\cite{Braaten:1993jw}
\bea
{\rm Re} \, \pi_{\rm T} & = & \omega_{\rm pl}^2 \left[ 1 + \frac{1}{2}G\left(v_*^2 k^2/\omega^2\right)\right]\, , \\
{\rm Re} \, \pi_{\rm L} & = & \omega_{\rm pl}^2\frac{m_{\gamma^\prime}^2}{\omega^2} \frac{ 1 - G\left(v_*^2 k^2/\omega^2\right)}{1-v_*^2 k^2/\omega^2} \,
\eea
for the plasma frequency 
\bea \omega_{\rm pl}^2 = \frac{4\pi\alpha n_e}{\sqrt{m_e^2 +(3\pi^2 n_e)^{2/3}}}
\eea with  the electron mass $m_e$ and the electron number density $n_e$. $v^*$ indicates the typical electron velocity, and $k = |\vec{k}|$.
The function $G$ is defined as
\bea
G\left(x\right) = \frac{3}{x}\left(1-\frac{2x}{3} -\frac{1-x}{2\sqrt{x}} \ln\frac{1+\sqrt{x}}{1-\sqrt{x}}\right) \, .
\eea
The imaginary parts of $\pi_{\rm T,L}$~\cite{Weldon:1983jn} come from the absorption rate of the electromagnetic excitations with the momentum from the dispersion relation of the external dark gauge boson as $\omega^2 - k^2  = m_{\gamma^\prime}^2$~\cite{An:2013yfc,Chang:2016ntp}. They become relevant at the resonance ${\rm Re}\,\pi_{\rm T,L} = m_{\gamma^\prime}^2$.
In the case that electrons are highly degenerate (i.e., $p_{F} \gg T_{\rm c}$ for  the electron Fermi momentum $p_{F} \simeq (3\pi^2 n_e)^{1/3}$ and the core temperature  $T_{\rm c} \simeq 30\,{\rm MeV}$) and relativistic (i.e., $m_e \ll p_{F}$), the real parts of $\pi_{\rm T,L}$ approximate
\bea
{\rm Re}\,\pi_{\rm T} & \simeq & \frac{3}{2}\, \omega_{\rm pl}^2 \, , \\
 {\rm Re}\,\pi_{\rm L} & \simeq & 3 \, \omega_{\rm pl}^2 \frac{m_{\gamma^\prime}^2}{\omega^2} \ln \left({\rm min}\,\left[\frac{p_F}{m_e}\, ,  \frac{\omega}{m_{\gamma^\prime}}\right]\right)\, 
 \label{eq:PiL}
\eea
when $m_\gamma' \lesssim \omega$.
We confirm that the resonance has a minor impact on the effective couplings, primarily because of the similar magnitude of the imaginary part (for transverse modes~\cite{Shin:2021bvz}) or the failure to satisfy the resonance condition (for longitudinal modes).

We will investigate two well-motivated dark gauge boson scenarios, which are presented in the following list.
\begin{itemize}
\item In the gauged $ B-L$ model, which is anomaly-free in the presence of right-handed neutrinos, 
the combination of $c_p-c_n$ that breaks isospin symmetry (which is zero at the vacuum) is effectively generated due to the medium effect as discussed above. Without the kinetic mixing, we obtain
\bea
\left. g^\prime \left(\tilde{q}_p^\prime - \tilde{q}_n^\prime\right)\right|_{B-L} = g^\prime \frac{\pi_{\rm T,L}}{m_{\gamma^\prime}^2 - \pi_{\rm T,L}} \, .
\label{eq:BLEff}
\eea
We observe that this effective violation of isospin symmetry is found to be the same for $U(1)$ symmetries that involve the electron number charge, such as the gauged $L_e-L_{\mu,\tau}$ model.

\item  Another model is the so-called dark photon model, in which no dark $U(1)$ charge is assigned to the SM particles (i.e., $q_f^\prime = 0$), but there is a presence of kinetic mixing denoted by $\varepsilon$. The effective isovector charge can be expressed as follows:
\bea
\left. g^\prime \left(\tilde{q}_p^\prime - \tilde{q}_n^\prime\right)\right|_{\rm DP} = \varepsilon e \frac{m_{\gamma^\prime}^2}{m_{\gamma^\prime}^2- \pi_{\rm T,L}} \, .
\label{eq:DPEff}
\eea
\end{itemize}

So far, we have primarily focused on the effective Lagrangian involving pions and nucleons ($p, n$). However, before we conclude this section, it is worth discussing the potential effects of other baryons, particularly the $\Delta$ resonance, on the production of dark gauge bosons. In the case of axion modes, for instance, processes such as $\pi^- p \rightarrow a\,n$ \cite{Carenza:2020cis, Fischer:2021jfm} can be additionally mediated by $\Delta$ baryons, resulting in an enhanced interaction rate \cite{Vonk:2022tho}. 
Therefore, it is important to examine whether a similar enhancement associated with the $\Delta$ resonance also occurs in the context of dark gauge boson production.

As we have discussed in this section, the interactions involving the $\Delta$ field can be described by terms that respect certain symmetries \cite{Gellas:1998wx, Pascalutsa:2005ts, Pascalutsa:2005vq, Pascalutsa:2006up}. The coupling $\gamma^\prime$-$N$-$\Delta$ is particularly relevant for contributions to $\pi^- p\to \gamma' n$ through an intermediate $\Delta$ resonance. This coupling is derived from the effective chiral Lagrangian involving photons (for isovector components), which arises from plasma mixing or the aforementioned kinetic mixing. In the framework of chiral expansion, the leading-order $\gamma^\prime$-$N$-$\Delta$ coupling emerges at $\mathcal{O}(\omega^2/m_N^2)$ in the form of the magnetic dipole moment due to the underlying spin-$3/2$ gauge symmetry \cite{Gellas:1998wx, Pascalutsa:2005ts, Pascalutsa:2005vq, Pascalutsa:2006up}. In comparison, we note that the $a$-$N$-$\Delta$ couplings in axion models occur at $\mathcal{O}(\omega/m_N)$ as the neutral pion field is replaced with the axion \cite{Tang:1996sq, Krebs:2009bf}.
Therefore, we do not expect a similar enhancement through the $\Delta$ baryon mediator in the case of dark gauge bosons, unlike what occurs in axion models. This justifies our simplification. 

\section{Dark gauge boson production from supernova pion abundance}
\label{sec:production}

\begin{figure}[h!]
\centering
\includegraphics[width=0.2\textwidth]{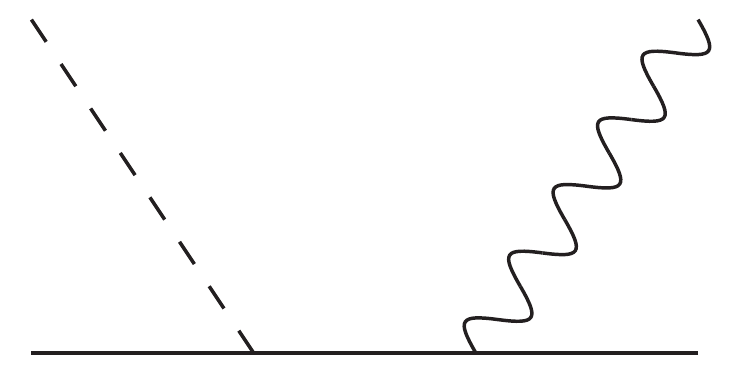}~~
\includegraphics[width=0.2\textwidth]{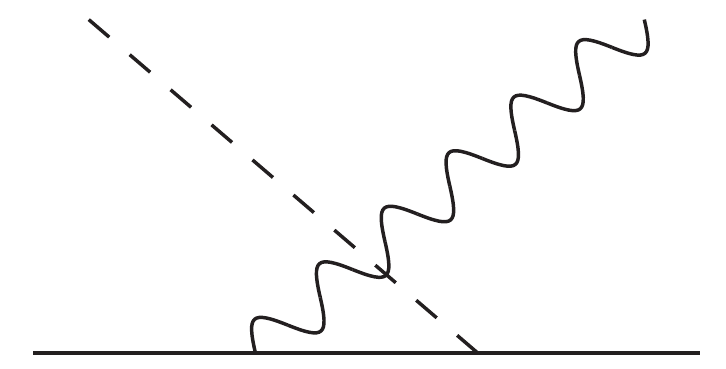}\\
\vspace{3.5mm}
\includegraphics[width=0.2\textwidth]{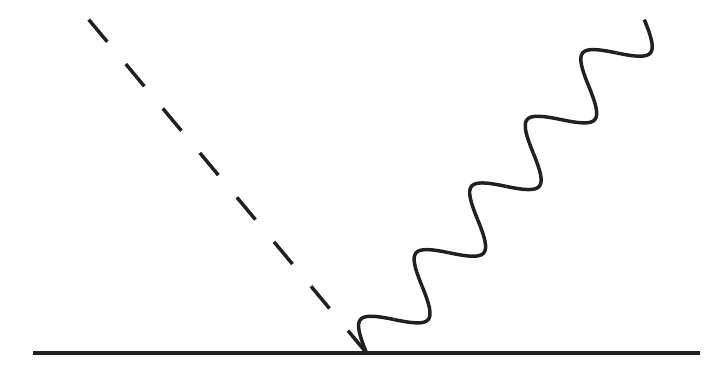}~~
\includegraphics[width=0.2\textwidth]{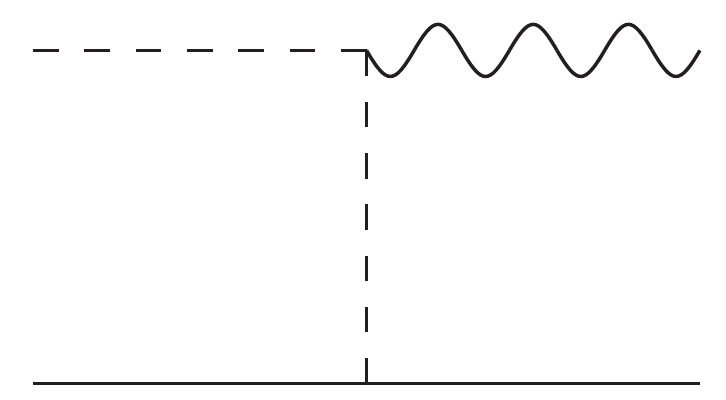}
\caption{\em The diagrams for the dark gauge boson (wavy) production via negatively charged pion (dashed) scatterings with nucleons (solid).}
\label{fig:diagrams}
\end{figure}

Let us now compute the rate of dark gauge boson production via pion-proton reactions, $\pi^- p \rightarrow \gamma^\prime n$.
The effective couplings
between the dark gauge boson and hadrons, as given in Eq.~\eqref{eq:VInt} leads to the four diagrams that are illustrated in Fig.~\ref{fig:diagrams}; the solid, dashed, and wavy lines represent the nucleon, pion, and dark gauge boson, respectively.
The external dark gauge boson can be attached to the nucleon line through the nucleon current couplings, which are depicted in the two upper diagrams. In addition, there is a four-point vertex originating from the contact  $\pi$-$N$-$\gamma^\prime$ interaction, represented by the lower left diagram. 
Finally,  the diagram mediated by the $\pi$-$\pi$-$\gamma^\prime$ couplings is shown in the lower right diagram.

Let us denote the four-momentum of the initial proton and the final neutron as $p_p$ and $p_n$, respectively, while the four-momentum of the incident negatively charged pion is represented by $p_\pi$.
The matrix elements with the effective couplings parametrized by $\tilde{q}_p$ and $\tilde{q}_n$ read as follows
\bea
\mathcal{M}_{N} & = &  \frac{\sqrt{2}}{f_\pi} \, g^\prime  \epsilon_\mu \, \bar{u}(p_n) \left[-\frac{\tilde{q}_n^\prime}{2p_n\cdot p_{\gamma^\prime}}\left(2 p_n^\mu \slashed{p}_\pi  + \gamma^\mu\slashed{p}_{\gamma^\prime}\slashed{p}_\pi  \right)\right.  \nonumber\\
&& \left. +\frac{\tilde{q}_p^\prime}{2p_p\cdot p_{\gamma^\prime}}\left(2 p_p^\mu \slashed{p}_\pi  - \gamma^\mu\slashed{p}_{\pi}\slashed{p}_{\gamma^\prime}  \right) \right]\gamma^5 u(p_p)  \, , \label{eq:MatrixN}\\
\mathcal{M}_{\rm con} & = &  \frac{\sqrt{2}}{f_\pi} \, g^\prime \left(\tilde{q}_n^\prime - \tilde{q}_p^\prime\right)\epsilon_\mu \, \bar{u}(p_n)\gamma^\mu \gamma^5 u(p_p)  \, , \label{eq:MatrixCon}\\
\mathcal{M}_{\pi} & = &   \frac{\sqrt{2}}{f_\pi} \,  g^\prime\left(\tilde{q}_n^\prime - \tilde{q}_p^\prime\right)   \bar{u}(p_n) \gamma^5 u(p_p)    \, \nonumber\\
&&\times \left[p_\pi^\mu\epsilon_\mu\right] \frac{2m_N}{p_\pi\cdot p_{\gamma^\prime} + \left(m_\pi^2-p_\pi^2\right)/2}\label{eq:MatrixPi}
\eea
 for the two upper diagrams, the lower left diagram, and the lower right diagram, respectively. Here, $u(p_N)$ denotes the spinor of a nucleon $N$, $m_N$ represents the nucleon mass, $m_\pi$ is the pion mass, and $\epsilon_\mu$ stands for the polarization vector of the dark gauge boson. In these expressions, as a reasonable approximation, we neglect the mass of the dark gauge boson, $m_{\gamma^\prime}$. 
At the vacuum where $p_\pi^2 = m_\pi^2$, one can straightforwardly 
verify the fulfillment of the Ward-Takahashi identity~\cite{Ward:1950xp,Takahashi:1957xn} (i.e., the current conservation law) by observing that the total amplitude $\mathcal{M}_{\rm tot} = \mathcal{M}_N + \mathcal{M}_{\rm con} + \mathcal{M}_\pi$ vanishes when replacing $\epsilon_\mu \rightarrow [p_{\gamma^\prime}]_{\mu}$.
This ensures that the four diagrams depicted in Fig.~\ref{fig:diagrams} constitute the complete set at the leading order.

The production rate of dark gauge bosons per unit volume can be described by
\bea
Q_{\gamma^\prime} & = &   \int \frac{d \vec{p}_p}{2E_p(2\pi)^3}\frac{d \vec{p}_\pi}{2E_\pi(2\pi)^3}\frac{d \vec{p}_n}{2E_n(2\pi)^3}\frac{d \vec{k}}{2\omega(2\pi)^3} \, \nonumber \\
&& \times \, \omega \left| \mathcal{M}_{\rm tot}\right|^2 f_p f_\pi \left(1-f_n\right) \nonumber \\
&&\times  \, \left(2\pi\right)^4\delta^{(4)} \left(p_p + p_\pi - p_n - p^\prime\right) \, ,
\label{eq:EmissivityEq}
\eea
where $f_N = \left(\exp\left[\left(E_N - \mu_N\right)/T\right] + 1\right)^{-1}$ represents the Fermi-Dirac distribution function for a nucleon $N$ with $\mu_N$ being its chemical potential. Here $E_i$ and $\vec{p}_i$ denote the energy and spatial momentum of each particle $i = (p,n,\pi)$.
For the pion occupation number, we approximate $f_\pi$ as $e^{- (E_\pi - \mu_\pi)/T}$, adopting the Boltzmann distribution instead of the Bose-Einstein distribution. This choice ensures consistency with the virial expansion, providing an appropriate approach to describing thermal pions in the presence of the nuclear medium~\cite{Fore:2019wib}. 

Before deriving the total scattering amplitude squared as the crucial ingredient in Eq.\eqref{eq:EmissivityEq}, we highlight some kinematic features of the Panofsky process $\pi^- p \rightarrow \gamma^\prime n$ \cite{Panofsky:1950he} that simplify the analytical evaluation related with the energy-momentum conservation as done in the axion case \cite{Carenza:2020cis,Fischer:2021jfm,Choi:2021ign}.
The integral over $d\vec{p}_n$ can be explicitly eliminated by utilizing the spatial momentum conservation, $\delta^{(3)}(\vec{p}_p +\vec{p}_\pi -\vec{p}_n -\vec{k})$.
Then, the energy conservation condition $\delta(E_p+E_\pi - E_n-\omega)$ reveals a subtle dependence on the angular relation between $\vec{k}$ and both $\vec{p}_p$ and $\vec{p}_\pi$. Nevertheless, due to the sufficiently low temperature and the chemical potential of nucleons in the pNS core, the nucleons remain massive and non-relativistic. This circumstance yields $E_n = E_p +\mathcal{O}(m_\pi^2 /m_N)$, allowing us to take $\omega\simeq E_\pi$ at a leading order approximation, i.e., $\delta(E_p+E_\pi - E_n-\omega)\simeq\delta \left(E_\pi - \omega\right)$. 
Therefore, the incident pion energy is efficiently converted into that of the emitted dark gauge boson through an elastic process.

For the vector current coupling, the spin sum for the dark gauge boson polarization vector $\epsilon_\mu$ is given by 
\bea
\sum_{\rm T} \epsilon_\mu^* \epsilon_\nu & \to & -g_{\mu\nu}  -\frac{m_{\gamma^\prime}^2}{k^2} g_{\mu 0}  g_{\nu 0} \, , \\
\sum_{\rm L} \epsilon_\mu^* \epsilon_\nu & \to & \frac{m_{\gamma^\prime}^2}{k^2} g_{\mu 0}  g_{\nu 0}
\label{eq:SumLong}
\eea
for the transverse ($\rm T$) and longitudinal ($\rm L$) components, respectively.	
In the high energy regime, we calculate the total squared scattering amplitude for each polarization taking the leading-order expansion with respect to $1/m_N$:
\bea
\left< \left| \mathcal{M}_{\rm tot}\right|^2_{\rm T} \right> & \simeq & g^{\prime 2} \left|\tilde{q}_n^\prime - \tilde{q}_p^\prime\right|^2 \frac{32 m_N^2}{f_\pi^2}  \nonumber \\
&& \times  \left[1+ \frac{p_\pi^2}{\tilde{\omega}^2}\left(1 - \frac{\tanh^{-1}\left[\left|\vec{p}_\pi \right| \omega /\tilde{\omega}^2\right]}{\left|\vec{p}_\pi \right| \omega / \tilde{\omega}^2}\right)\right] \, , ~~~ \label{eq:MtotT}\\
\left< \left| \mathcal{M}_{\rm tot}\right|^2_{\rm L} \right> & \simeq & g^{\prime 2} \left|\tilde{q}_n^\prime - \tilde{q}_p^\prime\right|^2 \frac{32 m_N^2}{f_\pi^2} \frac{m_{\gamma^\prime}^2}{\omega^2} \nonumber \\
&& \times \left[\left(\frac{\tanh^{-1}\left[\left|\vec{p}_\pi \right| \omega /\tilde{\omega}^2\right]}{\left|\vec{p}_\pi \right| \omega /\tilde{\omega}^2} -1\right) + \right. \nonumber\\
&& \left. +  \frac{p_\pi^2}{\tilde{\omega}^2}\left(\frac{\tanh^{-1}\left[\left|\vec{p}_\pi \right| \omega /\tilde{\omega}^2\right]}{\left|\vec{p}_\pi \right| \omega /\tilde{\omega}^2}-\frac{1}{2}\right)\right] \, , \label{eq:MtotL}
\eea
where the brackets denote an average over the angular dependence of $\hat{p}_\pi \cdot \hat{k}$ with the definition of $\hat{a} \equiv \vec{a}/|\vec{a}|$, and we define
\be
\tilde{\omega}  = \sqrt{\frac{\omega^2 + \left(\left| \vec{p}_\pi\right|^2 + m_\pi^2\right)}{2}} \, 
\label{eq:TildeOmega}
\ee
as the root mean square of the pion energy in the medium ($\omega$) and that in the vacuum ($ \sqrt{ \left| \vec{p}_\pi\right|^2 + m_\pi^2 }$).

In order to evaluate the dark gauge boson emissivity, 
we must determine the dispersion relation of nucleons and pions within a dense medium. To achieve this, we employ the phenomenological mean-field theory framework. Within this framework, nuclear interactions 
reduce the effective nucleon masses. At the non-relativistic regime, the nucleon energy can be expressed as~\cite{Martinez-Pinedo:2012eaj,Hempel:2014ssa,Carenza:2019pxu}
\bea
E_N  \simeq m_N + \frac{\left| \vec{p}_N \right|^2}{2m_N^*} + U_N \, ,
\eea
where $U_N$ is the mean-field potential energy~\cite{Skyrme:1959zz} obtained from Ref.~\cite{Fore:2019wib}, and $m_N^*$ denotes the Landau effective masses~\cite{Reddy:1997yr}. 
The ratio of $m_N^*$ to $m_N$ can be found in Ref.~\cite{Carenza:2019pxu}.

Recently, Ref.~\cite{Fore:2019wib}  has implemented the relativistic virial expansion as a model-independent approach to investigate the effects of the strong interactions between nucleons and pions on thermodynamic properties of pions. By utilizing the empirical s-wave and p-wave pion-nucleon phase shifts~\cite{Hoferichter:2015hva}, the enhanced number density of negative-charge pions is obtained.
Moreover, the pion self-energy is correlated with the pseudo-potential of the pion-nucleon interactions, which is also accounted for through the pion-nucleon phase shifts. These findings are reflected in the pion dispersion (for details, see Ref.~\cite{Fore:2019wib})
\be
E_\pi = \sqrt{\left| \vec{p}_\pi\right|^2 + m_\pi^2}+\Sigma_\pi \left(\left| \vec{p}_\pi\right|\right) \, ,
\label{eq:PionDispersion}
\ee
where $\Sigma_\pi\left(\left| \vec{p}_\pi\right|\right)$ is the real part of the momentum-dependent pion self-energy. The strength of the pseudo-potential, characterized by a fudge factor $\alpha$ in Ref.~\cite{Fore:2019wib}, is determined to match the thermal pion abundance in the virial expansion to that from the Boltzmann distribution with Eq.~(\ref{eq:PionDispersion}).
The modified pion dispersion relation may impact on core-collapse supernova dynamics, such as altering  the mean free path of low energy muon neutrinos as discussed in Ref.~\cite{Fore:2019wib}.

\begin{figure}[t!]
\centering
\includegraphics[width=0.45\textwidth]{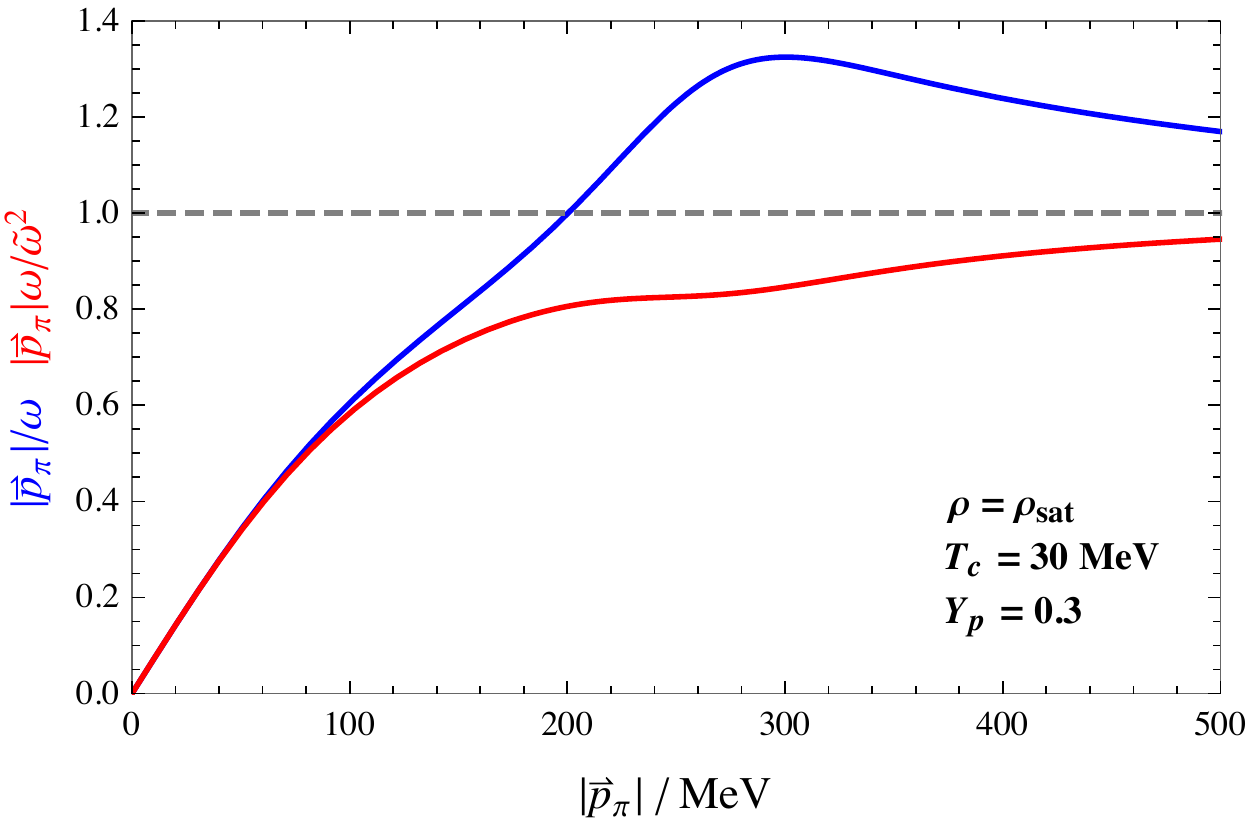}
\caption{\em $\left|\vec{p}_\pi\right|/\omega$ (blue) and $\left|\vec{p}_\pi\right|\omega/\tilde{\omega}$ (red) in function of the pion spatial momentum $\left|\vec{p}_\pi\right|$ at the fiducial pNS parameters of $\rho=\rho_{\rm sat}$, $T_c = 30\,{\rm MeV}$, and $Y_p = 0.3$. The dashed gray line indicates unity (i.e., light-like).}
\label{fig:PionDisp}
\end{figure}

Interestingly, the pion dispersion relation in Eq.~\eqref{eq:PionDispersion} exhibits a space-like behavior (i.e., $E_\pi < \left| \vec{p}_\pi\right|$) at rather high momentum values due to the strongly attractive p-wave interaction with nucleons~\cite{Fore:2019wib}.
In Fig.~\ref{fig:PionDisp}, we consider the pNS model parameters $\rho = \rho_{\rm sat}$, 
$T_c = 30\,{\rm MeV}$, and the proton fraction $Y_p \equiv n_p/n_B = 0.3$.
The blue line indicates $\left| \vec{p}_\pi \right| / \omega$, which indeed exceeds unity (dashed gray) at $\left| \vec{p}_\pi\right| > 200\,{\rm MeV}$.
This tachyonic behavior could be problematic when we compute the dark gauge boson emissivity because the phase space integral in Eq.~\eqref{eq:EmissivityEq} may contain a infrared divergence from the $t$-channel pion mediator in the lower right diagram in Fig.~\ref{fig:diagrams} with the matrix element given by Eq.~\eqref{eq:MatrixPi}.
To address this, we include the $(m_\pi^2-p_\pi^2)/2$ term in the denominator of Eq.~\eqref{eq:MatrixPi}, which acts as a regulator to mitigate the possible divergence arising from $p_\pi^2 \leq 0$ in the high-momentum regime.
The integration over the relative angle between $\hat{p}_\pi$ and $\hat{k}$ with the above prescription naturally reveals the argument of $\left| \vec{p}_\pi \right| \omega / \tilde{\omega}^2$ in the inverse hyperbolic tangent  functions in Eqs.~\eqref{eq:MtotT} and \eqref{eq:MtotL}, not $\left| \vec{p}_\pi \right| / \omega$ as expected in the vacuum. Such behavior is illustrated by the red line in Fig.~\ref{fig:PionDisp}.
In the graph, $\left| \vec{p}_\pi \right| \omega / \tilde{\omega}^2$asymptotically approaches but never crosses the dashed gray line, which represents unity.

We evaluate Eq.~\eqref{eq:EmissivityEq}  using the squared scattering amplitudes presented in  Eqs.~\eqref{eq:MtotT} and \eqref{eq:MtotL}. 
This allows us to approximate the volume emissivity of dark gauge bosons with the following equations 
\bea
\left[ Q_{\gamma^\prime}\right]_{\rm T} & \simeq &  \sqrt{\frac{2 m_N^{* 3} T^{11}}{\pi^{10}f_\pi^4}} z_p z_\pi \int  d x_p   \frac{x_p^2}{e^{x_p^2}+z_p}\frac{e^{x_p^2}}{e^{x_p^2} + z_n} \nonumber \\
&& \times  \int d x_\pi \left| g^{\prime }\left(\tilde{q}_n^\prime - \tilde{q}_p^\prime\right) \right|_{\rm T}^2 \,  \frac{x_\pi^2 \kappa_\pi}{e^{\kappa_\pi-y_\pi}} \nonumber \\
&&\times  \left[1+ \frac{\tilde{y}_\pi^2}{\tilde{\kappa}_\pi^2}\left(1 - \frac{\tanh^{-1}\left[x_\pi \kappa_\pi / \tilde{\kappa}_\pi^2\right]}{x_\pi \kappa_\pi / \tilde{\kappa}_\pi^2}\right)\right] \, , \\
\left[Q_{\gamma^\prime}\right]_{\rm L} & \simeq & \sqrt{\frac{2m_N^{* 3} m_{\gamma^\prime}^4T^{7}}{\pi^{10}f_\pi^4}} z_p z_\pi  \int  d x_p   \frac{x_p^2}{e^{x_p^2}+z_p}\frac{e^{x_p^2}}{e^{x_p^2} + z_n} \nonumber \\
&& \times  \int d x_\pi \left| g^{\prime } \left(\tilde{q}_n^\prime - \tilde{q}_p^\prime\right)\right|_{\rm L}^2 \, \frac{x_\pi^2 \kappa_\pi^{-1}}{e^{\kappa_\pi-y_\pi}} \nonumber\\
&& \times \Bigg[\left(\frac{\tanh^{-1}\left[x_\pi \kappa_\pi / \tilde{\kappa}_\pi^2\right]}{x_\pi \kappa_\pi / \tilde{\kappa}_\pi^2} -1 \right) + \nonumber \\
&& \quad + \frac{\tilde{y}_\pi^2}{\tilde{\kappa}_\pi^2}\left(\frac{\tanh^{-1}\left[x_\pi \kappa_\pi / \tilde{\kappa}_\pi^2\right]}{x_\pi \kappa_\pi / \tilde{\kappa}_\pi^2}-\frac{1}{2}\right)\Bigg] \, 
\eea
for the fugacity of a particle $i$, $z_i = \exp \left[\left(\mu_i - m_i\right)/T\right]$, $x_p = \left|\vec{p}_p\right|/\sqrt{2m_N^* T}$, $x_\pi = \left| \vec{p}_\pi\right|/T$, $\kappa_\pi = \omega/T$, $\tilde{\kappa}_\pi = \tilde{\omega}/T$, $y_\pi = m_\pi/T$, and $\tilde{y}_\pi^2 = p_\pi^2/T^2$.
It is important to note that the integration over $x_\pi$ must consider the momentum-dependent effective coupling $g^\prime (\tilde{q}_n-\tilde{q}_p)$ discussed in the previous section.

\begin{figure}[t!]
\centering
\includegraphics[width=0.45\textwidth]{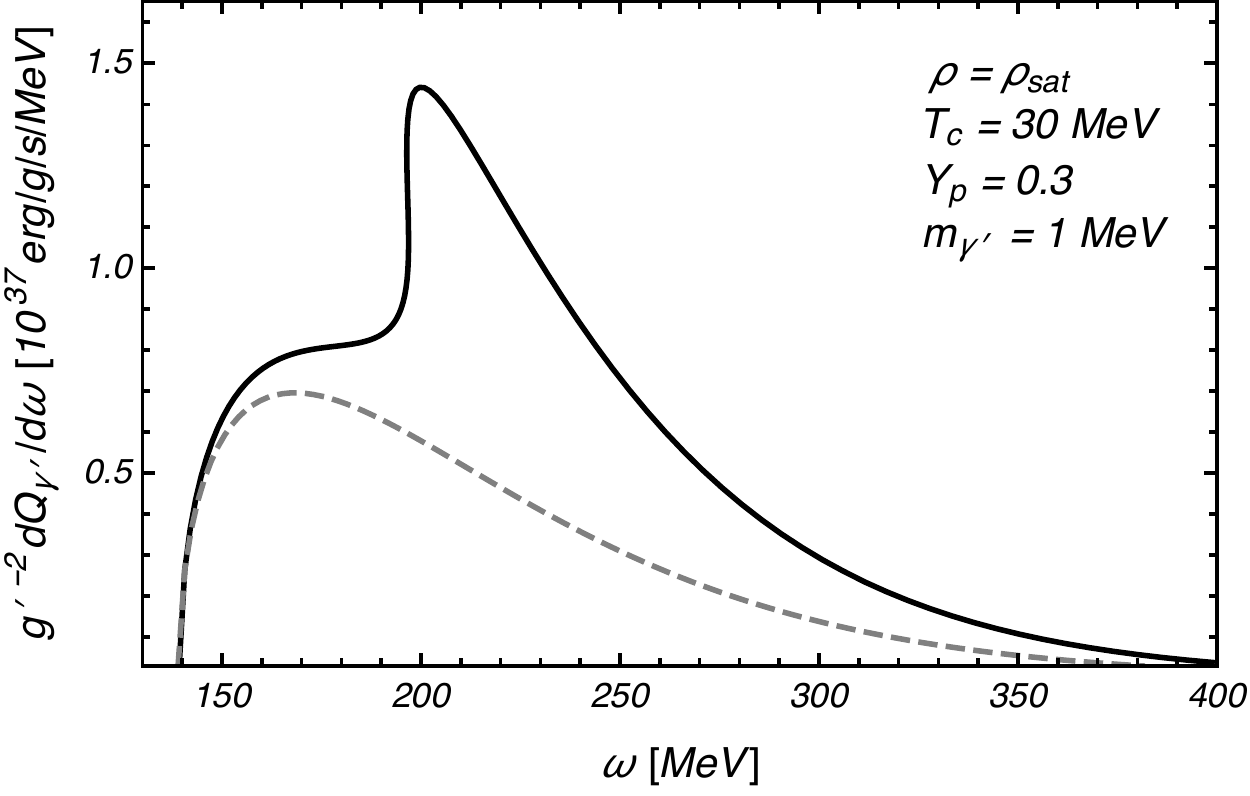}
\caption{\em The differential $\gamma^\prime$ emissivity via supernova pion reactions for the gauged $B-L$ model with $m_{\gamma^\prime} = 1\,{\rm MeV}$. The same pNS parameters in Fig.~\ref{fig:PionDisp}. The sold black and dashed gray lines take into account the $\pi^-$ dispersion relation in the pNS medium and the vacuum, respectively.}
\label{fig:Spectrum}
\end{figure}

We compute the dark gauge boson production rate $Q_{\gamma^\prime}$ numerically 
for the fiducial parameter set of $\rho = \rho_{\rm sat}$, $T_c= 30\,{\rm MeV}$, and $Y_p  = 0.3$ as the typical circumstance of the pNS interior at $1$ second after core bounce.
Following the approach of Ref.\cite{Fore:2019wib}, we determine the fugacities $z_p , z_n ,$ and $z_\pi$ using the mean-field theory formalism for nucleon-nucleon interactions and the virial expansion for pion-nucleon interactions. These methods account for the abundance of negative-charged pions $Y_{\pi^-} = \mathcal{O}(1)\%$ and the effective pion dispersion relation (Fig.~\ref{fig:PionDisp}).
In Fig.~\ref{fig:Spectrum}, the solid black line depicts the differential pion-induced dark gauge boson emissivity $dQ_{\gamma^\prime} /d\omega$ in the case of the gauged $B-L$ model with $m_{\gamma^\prime} = 1\,{\rm MeV}$. For the dark photon model, the overall spectrum is rescaled by replacing $g^\prime \pi_{\rm T}$ with $\varepsilon e m_{\gamma^\prime}^2$ since the transverse components dominate the total emissivity.
The numerical results indicates that the averaged dark gauge boson energy $\left<\omega\right>$ via pionic reactions approximates $200\,{\rm MeV}$. For comparison, we also include the dark gauge boson distribution assuming the $\pi^-$ dispersion relation in vacuum, represented by the dashed gray line. This shows clearly  a relatively suppression of the spectral function due to the smaller thermal occupation of $\pi^-$ abundance.

The noteworthy characteristic of pion-nucleon reactions is to generate dark gauge bosons in the energy range of $200\,{\rm MeV}$ to $300\,{\rm MeV}$. This energy range is approximately an order of magnitude larger than the core temperature which is typically expected in other thermodynamic processes such as nucleon-nucleon bremsstrahlung.
In the next section, we will discuss some phenomenological implications of dark gauge bosons associated with supernova.

\section{Implications}
\label{sec:implications}

Dark gauge bosons produced inside the pNS subsequently propagate outward.
They can be reabsorbed or trapped into the medium through its interactions with the SM particles.
These absorptive effects  may lead to various phenomenological consequences that will be explored in this section. 

There are several processes for dark gauge bosons to be reabsorbed or converted into SM particles in a medium, such as inverse bremsstrahlung ($\gamma^\prime NN \rightarrow NN$), inverse semi-Compton scattering ($\gamma^\prime e^- \rightarrow \gamma  e^-$), and decay (e.g., $\gamma^\prime \rightarrow e^- e^+$).
We observe that for masses above the $2m_e$ threshold, decays into electron-positron pairs typically dominate the absorptive width when dark gauge bosons couple to the electron current (as demonstrated in Ref.~\cite{Chang:2016ntp} for the dark photon model).  When considering decays of dark gauge bosons outside the pNS, where the electron density rapidly diminishes, the effects of Pauli blocking and plasma become negligible. Therefore, we can use the decay rate in vacuum given by:
\bea
\Gamma & = & \frac{g^{\prime 2}_e }{12\pi}\frac{m_{\gamma^\prime}^2}{\omega}\sqrt{1- \frac{4m_e^2}{m_{\gamma^\prime}^2}} \left(1+\frac{2m_e^2}{m_{\gamma^\prime}^2}\right)\, \Theta \left(m_{\gamma^\prime}-2m_e\right)  \,, ~~~~\label{eq:DecayT}
\eea
where $g_e^\prime = \left(-\varepsilon e + g^\prime q_e^\prime\right)$ denotes the coupling to the electron  in the vacuum.
In principle, the energy-dependent absorption rate has to be included in the integration over the produced dark gauge bosons spectrum.
However, as a good approximation for convenience, we simply estimate the decay rate at the typical energy of $\left<\omega\right> = 200\,{\rm MeV}$.

\subsection{Cooling argument}
\label{sec:constraints}

Observations of neutrinos from SN1987A by underground detectors (IMB~\cite{Bionta:1987qt}, Kamiokande II~\cite{Kamiokande-II:1987idp}, and Baksan~\cite{Alekseev:1987ej}) provide explicit evidence for the cooling of the pNS. Notably, the duration of the neutrino flux, which lasts for about 10 seconds after the bounce, is well explained by simulations assuming only the Standard Model, where the gravitational binding energy of the pNS is mainly released through energy diffusion at the neutrino sphere.

Novel particles can contribute to the cooling of the pNS, and if their cooling effect is significant, it could reduce the observed duration of the neutrino flux. However, to be consistent with the pNS evolution inferred from the neutrino observations, the net luminosity of novel particles streaming out of the pNS must be sub-dominant. This is a standard principle of the stellar cooling argument. The upper bound on the emissivity of novel particles is determined by the well-known Raffelt criterion established in Ref.~\cite{Raffelt:1996wa}. This criterion, given by
\begin{equation}
\frac{Q_{\gamma^\prime}}{\rho} e^{-\Gamma r_{\rm far}} < 10^{19}\,{\rm erg}\,{\rm g}^{-1}\,{\rm sec}^{-1}\,,
\label{eq:Raffelt}
\end{equation}
applies at 1 second after the bounce, assuming typical conditions of $\rho = \rho_{\rm sat}$ and $T=T_c$. Here, $r_{\rm far}$ represents the `far radius'~\cite{Chang:2016ntp}, which is introduced to measure the additional cooling in the heat reservoir (neutrino sphere) for supernova neutrinos. If a novel particle is absorbed within $r_{\rm far}$, its impact on the energy transfer that affects the observed neutrino signal is less significant. For our analysis, we approximately take $r_{\rm far} = 1000\,{\rm km}$ as a reasonable upper value based on the shock radius.

\subsection{Supernova explosion energy}
\label{sec:SNexplosion}

In addition to the extra energy leakage discussed above, novel particles propagating outside the pNS have the potential to transport their energy to the stellar envelope, thus contributing to the burst of the outer material (the supernova explosion).

The SN observables, such as light curves and spectroscopic ejecta expansion velocities, can be used to determine the explosion energy, as well as other explosion characteristics like the ejecta mass and the newly synthesized $^{56}{\rm Ni}$ mass, through numerical hydrodynamics simulations. Hydrodynamical modeling of core-collapse supernovae typically yields an explosion energy on the order of $\mathcal{O}(1)\,{\rm B}$, where ${\rm B}$ (bethe) is defined as $10^{51}\,{\rm erg}$. Therefore, the energy deposition in the stellar envelope resulting from the absorption of novel particles, as discussed in our analysis, should not exceed $2\,{\rm B}$, which corresponds to $1\,\%$ of the total binding energy of the pNS. This idea was initially suggested by Ref.~\cite{Falk:1978kf} and has been further explored in recent studies~\cite{Sung:2019xie,Caputo:2021rux,Caputo:2022mah}. We present roughly estimated constraints on dark gauge bosons given by
\bea
\frac{Q_{\gamma^\prime}}{\rho} \left( e^{-\Gamma r_{\rm far}} - e^{-\Gamma r_{\rm env}} \right)<  10^{17}\,{\rm erg}\,{\rm g}^{-1}\,{\rm sec}^{-1} , \, 
\label{eq:SNexp}
\eea
where $r_{\rm env}$ represents the radius of the progenitor star, assumed to be a red supergiant. In our analysis for type-II supernovae, we adopt $r_{\rm env} = 10^9\,{\rm km}$.

While the criterion in Eq.\eqref{eq:SNexp} is based on observations of typical explosion energies on the order of $\mathcal{O}(1)\,{\rm B}$, it can be refined by considering low-energy supernovae such as SN1054 leading to the Crab nebula \cite{Yang:2015ooa,Stockinger:2020hse} and many others~\cite{Pastorello:2003tc,Valenti_2009,Spiro_2014,Yang:2021fka}. These supernovae exhibit narrow spectral lines and lower luminosity at late times, indicating slower expansion velocities and lower ejected $^{56}{\rm Ni}$ masses (around $3$-$4$ times lower). As a result, the low-energy supernova explosion energy is on the sub-$\mathcal{O}(0.1)\,{\rm B}$, at least one order of magnitude below the typical value.

In general, novel particles are predominantly produced in the pNS, and their luminosity is less dependent on supernova modeling. Therefore, more stringent constraints can potentially be obtained from observations of low-energy supernovae. For further details and applications to axion physics, refer to Ref.~\cite{Caputo:2022mah}.

\subsection{Positron injection}
\label{sec:positron}

The diffuse gamma-ray emission from the Galactic center can be significantly enhanced by the injection of a large number of positrons originating from new particles. As these positrons traverse the interstellar medium, they interact with surrounding electrons through Coulomb interactions, resulting in radiation and finally  annihilation. The annihilation of positrons, primarily in the form of bound states with electrons (positronium), contributes to the observed $511\,{\rm keV}$ gamma-ray line~\cite{Churazov:2004as,Knodlseder:2005yq,Jean:2005af}. Precise measurements of the $511\,{\rm keV}$ line flux~\cite{Strong:2005zx,Bouchet:2010dj,Siegert:2015knp,Siegert:2019tus} using the SPI spectrometer~\cite{refId0} on the INTEGRAL satellite have placed conservative bounds on the galactic positron injection rate, limiting it to no more than $\mathcal{O}(10^{43})\,e^+\,{\rm s}^{-1}$~\cite{Prantzos:2010wi,DeRocco:2019njg}.

It is worth noting that the analysis of the measured gamma-ray spectrum at higher energy bands~\cite{Beacom:2005qv,Sizun:2006uh} may impose even more stringent constraints on the positron injection rate. Apart from annihilation at rest, propagating positrons also undergo in-flight annihilation, which contributes to the continuum flux above the $511\,{\rm keV}$ line. When positrons initially possess higher energy, their chances of surviving until they enter the non-relativistic regime decrease~\cite{Furlanetto:2002uq,Beacom:2005qv}, resulting in a larger flux of radiation above $511\,{\rm keV}$ over a wider energy range up to their initial energy~\cite{Sizun:2006uh}. For instance, $23\,\%$ of positrons with an initial energy of $100\,{\rm MeV}$ undergo in-flight annihilation, compared to $11\,\%$ for $10\,{\rm MeV}$ and $1.4\,\%$ for $1\,{\rm MeV}$. By considering mono-energetic positron injection above $100\,{\rm MeV}$ and the resulting radiation spectrum, the COMPTEL gamma-ray observations~\cite{Strong:1998ck} in the $1$-$3$, $3$-$10$, and $10$-$30\,{\rm MeV}$ energy bands have set an upper bound on the injection rate of approximately $\mathcal{O}(10^{42})\,e^+\,{\rm s}^{-1}$~\cite{Sizun:2006uh}.

Dark gauge bosons with masses $m_{\gamma^\prime}>2m_e$ could serve as a source of energetic positron injection in the galaxy~\cite{DeRocco:2019njg,Calore:2021lih}, provided that their decay length is sufficiently large compared to the escape radius $r_{\rm esc}$. Within $r_{\rm esc}$, interactions with the surrounding medium, such as annihilation with ambient electrons, significantly hinder the escape of positrons from the star. For type-II supernovae, $r_{\rm esc}$ is approximately $10^9\,{\rm km}$ during the relevant timescale before the shock wave reaches an outer layer~\cite{DeRocco:2019njg}, as it is equivalent to the radius of the progenitor star, $r_{\rm env}$.

Considering that type-II supernova events occur at a frequency of approximately 2 per century~\cite{Adams:2013ana}, the total amount of injected positrons from supernovae has to be less than $\mathcal{O}(10^{52})\,e^+$.
The resultant constraints can be written by
\bea
\left(\frac{Q_{\gamma^\prime}}{\rho} m_{\rm pNS}\right) \frac{\Delta t}{\left<\omega\right>}  e^{-\Gamma r_{\rm esc}} <  10^{52}\, e^+ \, 
\label{eq:PosInj}
\eea
with the time duration $\Delta t = 10\,{\rm sec}$.

\subsection{Absence of prompt $\gamma$-ray}
\label{sec:positron}

An additional constraint arises when considering SN1987A, particularly in scenarios where novel particles produced inside the proto-neutron star (pNS) decay outside the photosphere of the star. Within the photosphere, photons are effectively trapped and only emitted through thermal surface emission, making it difficult for highly energetic photons to directly escape the star. However, if novel particles are able to freely stream out of the star and subsequently decay into photons, we would expect gamma-ray signals to be detected at the same time as the neutrino observations, without any significant time delay.

At the time of the SN1987A neutrino observations, the gamma-ray spectrometer (GRS) on the Solar Maximum Mission satellite~\cite{1980SoPh} was in operation.
Since the detector was pointed toward the Sun at a $90^\circ$ angle relative to the direction of SN1987A, gamma-rays from SN1987A would bump into the side of the instrument, which was shielded by $2.5\,{\rm g}\,{\rm cm}^{-2}$ of spacecraft aluminium.
This shielding effect indirectly left signals in the detectors, allowing the estimation of the gamma-ray flux. However, no significant excess of gamma-ray signals above the expected background was detected by the GRS in the three energy bands of $4.1$-$6.4\,{\rm MeV}$, $10$-$25\,{\rm MeV}$, and $25$-$100\,{\rm MeV}$, for $223.2\,\sec$ before entering  calibration mode~\cite{Oberauer:1993yr}.
Consequently, an upper bound on the gamma-ray fluence, denoted as $\mathcal{F}_{\gamma}$, was determined for each energy band. The conservative limit, obtained by summing the fluences in all energy bins, is given as $\mathcal{F}_{\gamma} \lesssim 10\,{\rm cm}^{-2}$~\cite{DeRocco:2019njg}. In other words, the total number of escaped gamma-rays was constrained to be small as $N_{\gamma} < 4\times 10^{48}$.

The primary decay process of the leading dark gauge boson involving a photon is $\gamma^\prime \rightarrow e^- e^+ \gamma$. The partial decay width of this process is of $\mathcal{O}(\alpha)$ relative to that of the $e^-e^+$ channel.
See Appendix in Ref.~\cite{DeRocco:2019njg} for derivation of this partial decay width.
Consequently, we can derive the constraints based on the observations from the GRS as follows:
\bea
\left(\frac{Q_{\gamma^\prime}}{\rho} m_{\rm pNS}\right) \frac{\Delta t}{\left<\omega\right>}  e^{-\Gamma r_{\rm env}}  \,{\rm Br}\left( e^- e^+ \gamma \right) <  4 \times 10^{48} \, . \quad
\label{eq:promptGamma}
\eea 
Here, ${\rm Br}\left( e^- e^+ \gamma \right)$ represents the branching ratio of the $\gamma^\prime \rightarrow e^- e^+ \gamma$ channel.

\subsection{Results}
\label{sec:results}

\begin{figure}[t!]
\centering
\includegraphics[width=0.45\textwidth]{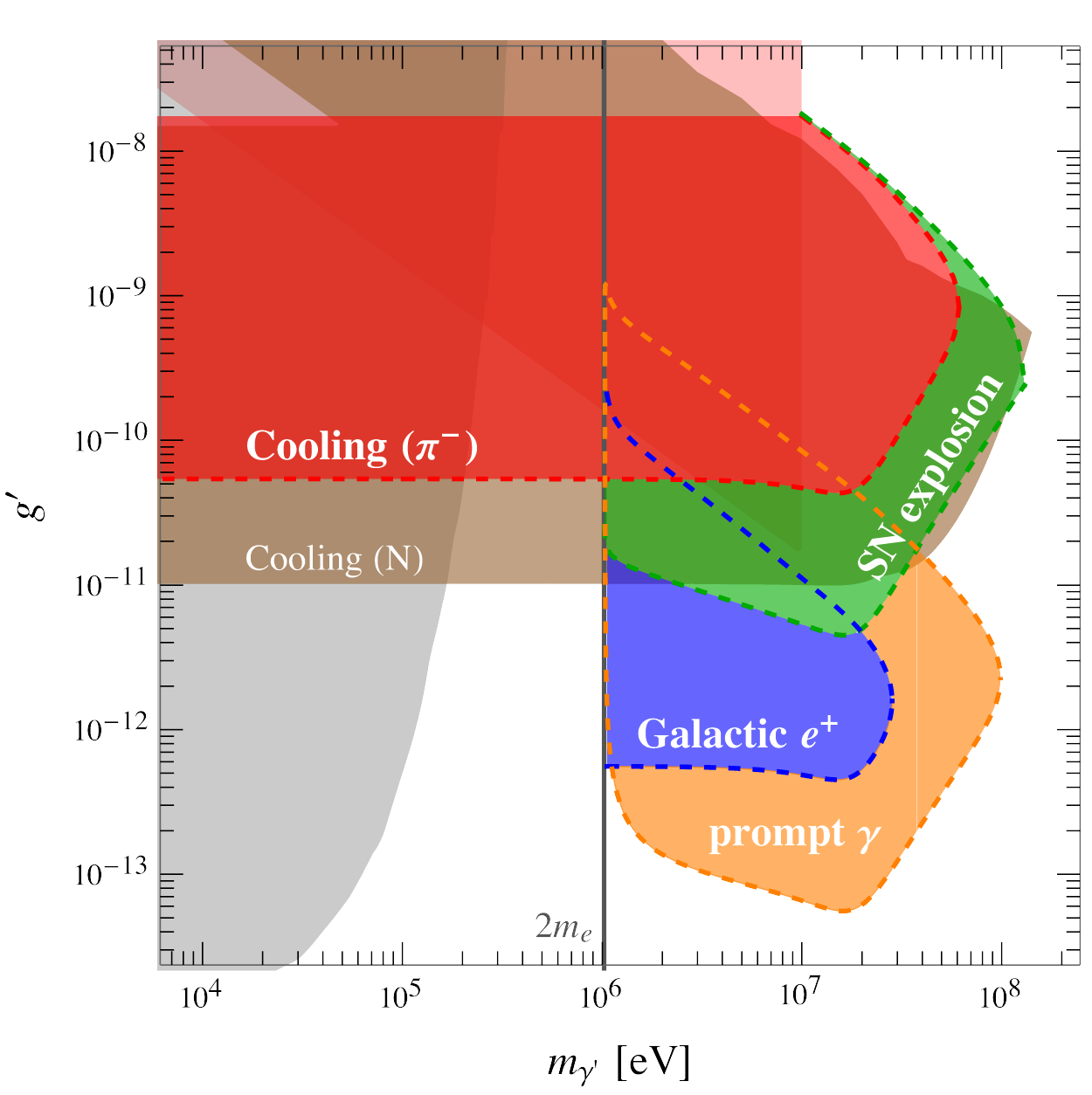}
\caption{\em The constraints plot for the gauged $B-L$ model. The red, green, blue, and orange regions are excluded by the cooling argument on SN1987A, the observed supernova explosion energy, the galactic positron injection, and non-observations of gamma-ray excess in SN1987A, respectively. The brown region corresponds to the previous SN1987A cooling bound in terms of nucleon-nucleon bremsstrahlung radiation~\cite{Shin:2021bvz}. The gray and pink regions come from the cooling argument on the red giants~\cite{An:2014twa,Hardy:2016kme}, and neutron stars~\cite{Hong:2020bxo}, and from Big Bang nucleosynthesis~\cite{Heeck:2014zfa,Knapen:2017xzo}, respectively.} 
\label{fig:BLconstraints}
\end{figure}

Fig.~\ref{fig:BLconstraints} illustrates the excluded parameter region of the gauged $B-L$ model.
The constraints with the dashed boundary are newly excluded by the arguments discussed in this paper.
The red region corresponds to the stellar cooling constraint based on the production of gauged $B-L$ bosons through thermal pion scatterings, as described in Sec.~\ref{sec:production}.
In comparison with the brown-shaded region for the preceding SN1987A cooling constraint, which
considers bremsstrahlung of dark gauge bosons via nucleon-nucleon scatterings~\cite{Shin:2021bvz}  (notice that the estimation is performed within a more practical approach of the pNS profile), the pion-induced cooling bound is less stringent due to the somewhat less efficient energy loss rate.

For masses above $2m_e$, the constraint on the supernova explosion energy given in Eq.~\eqref{eq:SNexp} corresponds to the green region, which is comparable to the cooling argument. Furthermore, the contributions to galactic positron injection and a prompt SN1987A gamma-ray signal impose significant constraints on the blue and orange regions, respectively.
The gray and pink regions represent preexisting constraints from the cooling argument on low-density stars~\cite{An:2014twa,Hardy:2016kme}, young isolated neutron stars (such as those in Cas A and SN1987A)\cite{Hong:2020bxo}, and Big Bang nucleosynthesis\cite{Heeck:2014zfa,Knapen:2017xzo}, respectively.

\begin{figure}[t!]
\centering
\includegraphics[width=0.45\textwidth]{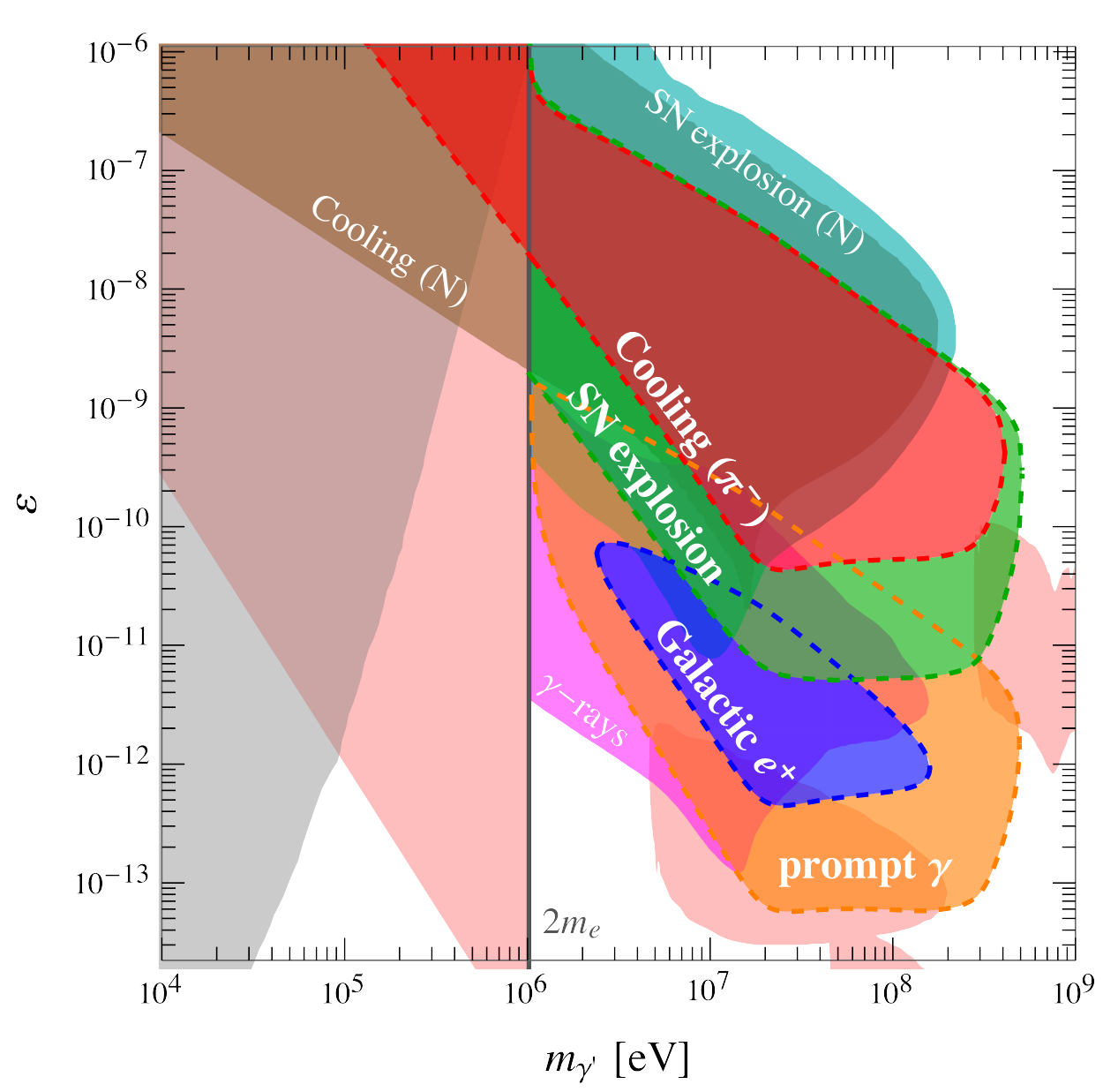}
\caption{\em The constraints plot for the dark photon model. The red, green, blue, and orange regions are excluded by the cooling argument on SN1987A, the observed supernova explosion energy, the galactic positron injection, and non-observations of gamma-ray excess from SN1987A, respectively. The brown, cyan, and magenta regions represent the previous bounds from the SN1987A neutrino observations~\cite{Chang:2016ntp}, the supernova explosion energy~\cite{Sung:2019xie}, and the astrophysical gamma-ray signals~\cite{DeRocco:2019njg,Calore:2021lih}, respectively. The other current bounds from the stellar evolution~\cite{An:2013yfc,Redondo:2013lna} (gray) and cosmology~\cite{Redondo:2008ec,Prantzos:2010wi,Fradette:2014sza} (pink).}
\label{fig:DPconstraints}
\end{figure}

Figure~\ref{fig:DPconstraints} presents the exclusion plot for the dark photon model. Each colored region, delineated by a dashed boundary, is disfavored due to phenomenological considerations related to dark photon production inside the pNS through scatterings involving $\pi^-$, as illustrated in Fig.~\ref{fig:BLconstraints}.
The constraints on the dark photon in the previous literature are given by 
the brown, cyan, and magenta region from the observed neutrino cooling of SN1987A~\cite{Chang:2016ntp}, the explosion energy~\cite{Sung:2019xie}, and the gamma-ray observations associated with core-collapse supernovae~\cite{DeRocco:2019njg}, respectively. These constraints overlook the contribution of thermal pions. 
The current bounds based on the stellar evolution~\cite{An:2013yfc,Redondo:2013lna} and Big Bang nucleosynthesis~\cite{Redondo:2008ec,Prantzos:2010wi,Fradette:2014sza} are depicted as the gray and pink regions, respectively.

It is worth noting that the constraints associated with thermal pions (indicated by the regions with dashed line boundaries in Figs.\ref{fig:BLconstraints} and \ref{fig:DPconstraints}) are primarily determined by the transverse components of dark gauge bosons. The characteristic of $\omega \gg \omega_{\rm pl}\, , T$ for pionic processes accounts for a weaker emissivity of the longitudinal polarization by the factor of $(m_{\gamma^\prime}^2/\omega^2)$ in Eq.~\eqref{eq:SumLong} from the current conservation law and/or $(\omega_{\rm pl}^4/\omega^4)$ from the plasma effect on the couplings of longitudinal gauged $B-L$ bosons (see the effective isospin breaking term in Eq.~\eqref{eq:BLEff} with Eq.~\eqref{eq:PiL} for $\pi_{\rm L}$).  Even if a sizable coupling strength could generate a sufficient amount of longitudinal dark gauge bosons to leave an imprint on the observables, such dark gauge bosons are immediately reabsorbed into the medium near their production point that leads to an exponentially suppressed net luminosity.

The effective isovector terms $g^\prime (\tilde{q}_p^\prime-\tilde{q}_n^\prime)$ of gauged $B-L$ bosons and dark photons have the distinct feature as described in Eq.~\eqref{eq:BLEff} and Eq.~\eqref{eq:DPEff}, respectively, depending on the  mass values of the dark gauge bosons.
The behavior of the emissivity for the transverse components, with ${\rm Re}\, \pi_{\rm T} = \mathcal{O}(1)\, \omega_{\rm pl}^2$ being constant over the momentum space of interest, relies on the competition between the dark gauge boson mass $m_{\gamma^\prime}$ and the plasma frequency $\omega_{\rm pl}\sim 10\,{\rm MeV}$.
\begin{itemize}
\item In the regime of $m_{\gamma^\prime}^2 \ll \omega_{\rm pl}^2$, we find $(\tilde{q}_p^\prime-\tilde{q}_n^\prime) \propto m_{\gamma^\prime}^0$ for the gauged $B-L$ model and $ (\tilde{q}_p^\prime-\tilde{q}_n^\prime) \propto m_{\gamma^\prime}^2$ for the dark photon model.
Thus, the upper bounds on dark photons is quadratically weaker for a lower mass, whereas those on gauged $B-L$ bosons have flat (the red and blue in Fig.~\ref{fig:BLconstraints}) or a mild mass dependence due to  the decay length being comparable to the size of the progenitor and the branching ratio of the three body decay $\gamma^\prime \rightarrow e^- e^+ \gamma$ (the green and orange  in Fig.~\ref{fig:BLconstraints}).

\item In the regime where the dark gauge boson masses are above the plasma frequency, $m_{\gamma^\prime}^2 \gg \omega_{\rm pl}^2$, the effective coupling behaviors change. For the gauged $B-L$ model, we have $(\tilde{q}_p^\prime-\tilde{q}_n^\prime) \propto m_{\gamma^\prime}^{-2}$, while for the dark photon model, $(\tilde{q}_p^\prime-\tilde{q}_n^\prime) \propto m_{\gamma^\prime}^0$.
In contrast to the low-mass limit discussed above, the upper bounds on gauged $B-L$ bosons are now quadratically weaker for a larger mass, and those on dark photons become relatively constant  up to around $\left<\omega\right>\approx  200\,{\rm MeV}$.
\end{itemize}
We observe that the lower bounds are strongly influenced by the decay length, as described in Eq.~\eqref{eq:DecayT}, resulting in inverse linear slopes considering  $\Gamma_{\gamma^\prime} \propto \left(g^\prime m_{\gamma^\prime}\right)^2$.
As a result, the constraints based on pionic processes do not reach the mass ranges above $100\,{\rm MeV}$ in the gauged $B-L$ model, whereas they are attainable in the dark photon model.

\section{Conclusions and discussion}
\label{sec:conclusions}

We have extensively explored the production of dark gauge bosons induced by pions in the core-collapse supernova,  particularly in the formation of protoneutron stars (pNS) with the high temperature $T\sim 30\,{\rm MeV}$ and the large electron chemical potential $\mu_e = \mu_n - \mu_p \sim 200\,{\rm MeV}$. 
As discussed in Sec.~\ref{sec:production}, the presence of negatively charged pions in a hot and dense environment plays a significant role for the dark gauge boson production and cannot be disregarded. In the framework of proper effective theories for an additional $U(1)$ gauge boson, described in Sec.~\ref{sec:effective}, the symmetry argument in ChPT accounts  for the isospin breaking aspect of pionic processes that are also subject to the plasma effects.
This leads to the couplings between the dark gauge boson to the electromagnetic current: dark photons through the kinetic mixing, and gauged $B-L$ bosons through the plasma mixing~\cite{Rrapaj:2015wgs}.
To determine the emissivity of dark gauge bosons, we conducted both analytical and numerical calculations, utilizing the reliable method of the relativistic virial expansion proposed in Ref.~\cite{Fore:2019wib}. 
This effectively characterizes the strongly interacting pion-nucleon plasma.

We have discussed the pion-induced dark gauge boson production in a core-collapse supernova, which is forming the pSN with the high temperature $T\sim 30\,{\rm MeV}$ and the large electron chemical potential $\mu_e = \mu_n - \mu_p \sim 200\,{\rm MeV}$.
It is enough for negatively charged pions to comprise a non-negligible fraction in a hot dense matter (Sec.~\ref{sec:production}).
In the context of the proper effective theory frameworks for an additional $U(1)$ gauge boson (Sec.~\ref{sec:effective}), the symmetry argument in ChPT accounts for the isospin breaking aspect of pionic processes that could be affected by the plasma effects.
It connects dark gauge bosons to the electromagnetic current; dark photons through the kinetic mixing and gauged $B-L$ bosons through the plasma mixing~\cite{Rrapaj:2015wgs}.
We computed the emissivity of dark gauge bosons analytically and numerically within the reliable method of the relativistic virial expansion~\cite{Fore:2019wib} that describes the strongly-interacting pion-nucleon plasma .

The produced dark gauge bosons can have an impact on various observations associated with supernovae.
We have examined several phenomenological implications of dark gauge bosons (Sec.~\ref{sec:implications}): the observed neutrino burst of SN1987A, the inferred supernova explosion energy, the galactic positron injection to cause diffuse gamma-ray emissions from the Galactic center, and the non-observation of prompt gamma-rays in SN1987A. The implications depend on the distribution of energy carried away or accumulated in specific regions of the star by dark gauge bosons. In models where dark gauge bosons couple to the electron current, such as the dark photon and the gauged $B-L$ models, the dominant contribution to the optical depth of a dark gauge boson with a mass greater than $2m_e$ arises from its decay into electron-positron pairs. Each observational consequence gives rise to a corresponding bound, which can be estimated analytically with an order of magnitude approximation, as expressed in Eqs~\eqref{eq:Raffelt}-\eqref{eq:promptGamma}. These constraints allow us to establish limits on the gauged $B-L$ and dark photon models, which are illustrated in Fig.~\ref{fig:BLconstraints} and Fig.~\ref{fig:DPconstraints}, respectively.

It is noticed that the newly derived constraints in this work for the gauged $B-L$ model and the dark photon model can be applied to other scenarios as well. 
The isovector structures described in Eq.~\eqref{eq:BLEff} and Eq.~\eqref{eq:DPEff} for the gauged $B-L$ model and the dark photon model require a charge assignment of $q_p^\prime - q_n^\prime = 0$ and $q_p^\prime - q_n^\prime + q_e^\prime= 0$ (or equivalently a nonvanishing kinetic mixing), respectively.
In addition, the condition $q^\prime_e \neq 0$ is a commone requirement for decay into electron-positron pairs.
For example, our results can be utilized to constrain the lepton flavor-dependent gauged $U(1)$ extensions of the Standard Model~\cite{Chun:2022qcg}. 
In the case of gauged $L_e - L_{\mu / \tau}$ models, which satisfy  $q^\prime_e\neq 0$ and $q^\prime_p-q^\prime_n=0$, the constraints derived from supernova pions explicitly follow the results in Fig.~\ref{fig:BLconstraints} for the gauged $B-L$ model. 
In the gauged $L_\mu - L_\tau$ model, despite the absence of tree-level $q_e^\prime$, the irreducible kinetic mixing from the muon and tau loop imposes the constraints that can be inferred from Fig.~\ref{fig:DPconstraints} for the dark photon model, taking into account the the loop-suppressed rescaling factor.

We have considered energy transfer of dark gauge boson to visible particles through the its decays such as $\gamma^\prime \rightarrow e^-e^+$, $\gamma^\prime \rightarrow e^- e^+ \gamma$, and $\gamma^\prime \rightarrow \gamma\gamma \gamma$. However, even if $m_{\gamma^\prime} < 2m_e$, the energy of dark gauge bosons could be deposited in a medium or converted into the creation of SM particles.
Within the star, the interaction between dark gauge bosons and matter causes their kinetic energy to be redistributed within the medium. The efficiency of this kinematic trapping or reabsorption depends on the density of the medium, which decreases sharply outside the pNS.
Consequently, the energy transport only affects the cooling of the pNS by dark gauge bosons. Roughly speaking, considering the lower emissivity from pionic processes and a similar optical length compared to nucleon-nucleon bremsstrahlung~\cite{Shin:2021bvz}, we do not expect the lower bound from pion-induced cooling to differ significantly from that of nucleon-nucleon bremsstrahlung. This explains why the top boundary of the red region in Fig.~\ref{fig:BLconstraints} is artificially truncated.

In addition to the decay of dark gauge bosons, they can also undergo conversion into photons. This conversion arises from the mixing between dark gauge bosons and photons, which is induced by either kinetic mixing parametrized by $\varepsilon$ or plasma effects (as discussed around Eq.~(10) in \cite{Hong:2020bxo}). Generally, the probability of conversion during propagation is typically on the order of $\mathcal{O}(g^{\prime 2})$, representing the square of the mixing angle. However, at resonance points where $m_{\gamma^\prime}^2 = \pi_{\rm T, L}$, the conversion probability can be dramatically enhanced. This resonant conversion of dark gauge bosons to photons outside the star can result in an instantaneous gamma-ray flux. However, the energy of the converted photons corresponds to that of the dark gauge bosons, which is above the energy range measured by the Gamma-Ray Spectrometer (GRS).  On the other hand, transitions occurring inside the star would contribute to the supernova explosion. The rate of resonant conversion is determined by the adiabaticity at the resonance point, which depends on how the quantity $\pi_{\rm T, L} \propto n_e$ varies in space. 
In the envelope regime of the progenitor, where $ {\cal O}(10^4)\,{\rm km} \lesssim r \lesssim  r_{\rm env} $, the electron density rapidly decreases as we move outward. Consequently, the substantial non-adiabaticity results in a small conversion rate. However, when considering the outer region of the progenitor core spanning from approximately $r_{\rm far}=\mathcal{O}(10^3)\,{\rm km}$ to $\mathcal{O}(10^4)\,{\rm km}$, where the shock wave has not yet reached, the density appears to remain relatively constant at around $\mathcal{O}(10^6)\,{\rm g}\,{\rm cm}^{-3}$.
Therefore, within this regime, transverse dark gauge bosons with a mass of $\mathcal{O}(10)\,{\rm keV}$ may efficiently convert into photons through an adiabatic level-crossing transition. Nevertheless, conducting a detailed investigation of the progenitor profile is imperative, as it goes beyond the scope of this paper. We leave the comprehensive discussion and analysis of this matter to future work.

Finally, we would like to address important uncertainties regarding our analysis. The concise analytic expressions presented in Eqs.~\eqref{eq:Raffelt}-\eqref{eq:promptGamma} are based on a simplified assumption that the interior of the pNS is homogeneous in terms of thermodynamic properties. This single-zone approximation may oversimplify the actual pNS profile, particularly when considering thermal pions.
The thermodynamic variables we have considered ($\rho=\rho_{\rm sat}$, $T=30\,{\rm MeV}$, and $Y_p = 0.3$) correspond to typical conditions at the surface of the pNS, approximately 1 second after core bounce. Due to the relatively high temperature, thermal negatively-charged pions mainly reside near the surface, as described in Ref.~\cite{Fischer:2021jfm}. Therefore, a more realistic pNS profile would likely lead to an $\mathcal{O}(1)$ modification to the pionic dark gauge boson emissivity, resulting in minor changes to the constraints shown in Fig.~\ref{fig:BLconstraints} and \ref{fig:DPconstraints}. For further insights into the single-zone approximation in the context of the supernova explosion argument, we refer to Ref.~\cite{Caputo:2022mah}.
It is important to note that the results based on Eqs.~\eqref{eq:Raffelt}-\eqref{eq:promptGamma} should be considered as rough estimates. 
Inclusion of bremsstrahlung dark gauge bosons through nucleon scatterings~\cite{Shin:2021bvz} for each implication in Sec.~\ref{sec:implications} could enhance our results. However, as pointed out in Ref.~\cite{Carenza:2019pxu} regarding the axion case, estimating nucleon-nucleon bremsstrahlung scatterings involves theoretical uncertainties, potentially leading to rates that are orders of magnitude lower than those computed using the first-order approximation, such as the one-pion-exchange approximation.
On the other hand, our investigation into the production of dark gauge bosons via pion-induced scatterings, as extensively explored in this paper, reveals its robustness. This is attributed to our utilization of a model-independent methodology, the relativistic virial expansion~\cite{Fore:2019wib}, which relies on empirical information regarding pion-nucleon scattering and its corresponding phase shift~\cite{Hoferichter:2015hva}.

\begin{acknowledgments}
The authors acknowledge Doojin Kim, Pyungwon Ko, and Giuseppe Lucente for useful discussions. This work is supported by the research grant: “The Dark Universe: A Synergic Multi-messenger Approach” number 2017X7X85K under the program PRIN 2017 funded by the Ministero dell’Istruzione, Universita` e della Ricerca (MIUR); “New Theoretical Tools for Axion Cosmology” under the Supporting TAlent in ReSearch@University of Padova (STARS@UNIPD) and by IBS under the project code, IBS-R018-D1. S.Y. is supported by Istituto Nazionale di Fisica Nucleare (INFN) through the Theoretical Astroparticle Physics (TAsP) project. C.S.S. acknowledges support from the National Research Foundation of Korea (NRF-2022R1C1C1011840 and NRF-2022R1A4A5030362). S.Y. would like to thank Deutsches Elektronen-Synchrotron (DESY) for the kind hospitality during this work was in progress.
\end{acknowledgments}


\bibliographystyle{apsrev4-2}
\bibliography{PionDarkVector}

\end{document}